\documentclass[aps,prb,10pt,amsmath,twocolumn,amssymb,superscriptaddress,showpacs]{revtex4-1} 
\usepackage{bm}
\usepackage[]{graphicx}
\usepackage[tight]{units}
\usepackage{color}
\usepackage{ulem}
\usepackage{marvosym}
\usepackage{mathtools}
\usepackage{afterpage}
\usepackage[colorlinks=true,citecolor=blue,linkcolor= blue,linkcolor=blue,urlcolor=blue]{hyperref}

\DeclareMathAlphabet\mathbfcal{OMS}{cmsy}{b}{n}
\usepackage{calrsfs}
\makeatletter
\def\mathcolor#1#{\@mathcolor{#1}}
\def\@mathcolor#1#2#3{%
  \protect\leavevmode
  \begingroup
    \color#1{#2}#3%
  \endgroup
}
\makeatother

\newcommand{\F}[1]{Fig.~\ref{#1}}

\newcommand{\Eq}[1]{Eq.~\ref{#1}}
\newcommand{\eq}[1]{eq.~\ref{#1}}

\newcommand{\ko}{\mathbf{k}}
\newcommand{\kp}{\mathbf{k}+\bm{q}}

\newcommand{\q}{\mathbf{q}\nu}

\newcommand{\E}{\epsilon_{\ko}^{0}}
\newcommand{\Ep}{\epsilon_{\kp}^{0}}

\newcommand{\me}{\mathfrak{g}_{\ko}^{\q}}

\newcommand{\delpmE}{\delta\left(E-\Ep \pm \hbar\omega_{\q}\right)}

\begin{document}
\title{Electron-phonon interaction and transport properties of  \\ metallic bulk and monolayer transition metal dichalcogenide TaS$_2$}
\author{Nicki Frank Hinsche}
\email{nickih@fysik.dtu.dk}
\author{Kristian Sommer Thygesen}
\affiliation{Center for Atomic-scale Materials Design, Department of Physics, Technical University of Denmark, DK-2800 Kgs. Lyngby, Denmark}

\date{\today}

\begin{abstract}
Transition metal dichalcogenides have recently emerged as promising two-dimensional materials 
with intriguing electronic properties. Existing calculations of intrinsic phonon-limited electronic transport so far 
have concentrated on the semicondcucting members of this family. 
In this paper we extend these studies by investigating the influence of electron-phonon coupling 
on the electronic transport properties and band renormalization of prototype inherent metallic bulk and monolayer TaS$_2$. 
Based on density functional perturbation theory and semi-classical Boltzmann transport calculations, 
promising room temperature mobilities and sheet conductances are found, which can compete with other established 2D materials, leaving TaS$_2$ as promising material candidate for transparent conductors or as atomically thin interconnects. Throughout the paper, the electronic and transport properties of TaS$_2$ are compared to those of its isoelectronic counterpart TaSe$_2$ and additional informations to the latter are given. 
We furthermore comment on the conventional superconductivity in TaS$_2$, where no 
phonon-mediated enhancement of $T_C$ in the monolayer compared to the bulk state was found. 
\end{abstract}

\maketitle
\subsection{Introduction}
Succeeding the many advances in fundamental science and applications of graphene, other two-dimensional materials, especially the transition-metal dichalcogenides (TMDs), have attracted remarkable interest for their appealing electrical, optical and mechanical properties. 
TMDs form layered compounds having a metal layer sandwiched between two chalcogen layers. These monolayers are weakly bound to each other by the dispersive van der Waals interaction, leading to a quasi-two-dimensional behavior in the monolayers \cite{Kuc:2015da,Kolobov:2016uw}. 
Among the TMDs, exfoliated semi-conducting materials, e.g. MoS$_2$ and WS$_2$, have been extensively in focus for their promising electronic mobilities and thus a high potential in thin-film transistor applications \cite{Kim:2012ft,Kaasbjerg:2012uy,Ovchinnikov:2014kv}. 
Recently metallic TMDs are in focus for their potential as transparent conductive electrodes and gained increasing importance for information-, optoelectronic- and energy-applications \cite{Ellmer:2012hm,Gjerding:2017ee}. 
Crucial requirements for these are high electrical conductivity and high transparency; properties that are often contradicting. 

Based on density functional theory and semi-classical Boltzmann transport, we will elaborate and discuss the influence of electron-phonon coupling onto the electronic transport properties and band renormalization of prototype metallic bulk and monolayer TaS$_2$ and compare it to conventionally used two-dimensional and bulk materials. Additional details on to the isoelectronic 
counterpart monolayer TaSe$_2$ are given along the discussion and can be found in the supplemental material~\cite{supp}. 
Furthermore the influence of low-dimensionality on the phonon-mediated superconductivity in metallic TaS$_2$ will be discussed. While the Mermin-Wagner theorem \cite{Mermin:1966da,Hohenberg:1967br,Koma:1992dg} suggests that in two-dimensional systems no superconductivity, or even an enhancement compared to its bulk state, should be observed, opposite trends have been discovered recently, including metallic TaS$_2$ \cite{Zhang:2010ja,Noffsinger:2010iv,NavarroMoratalla:2016df,Uchihashi:2017ig}.


\subsection{Methodology}
The electronic and phonon properties of monolayer and bulk TaS$_2$ were calculated within the framework of fully-relativistic density functional perturbation theory as implemented in a modified version of the Quantum Espresso suite \cite{Giannozzi:2009hx,Zahn:2011ci,Rittweger:2016vo}. 
We focus on the H-polytypes, having a hexagonal Bravais lattice with undistorted trigonal prismatic coordination. To account for the weakly bonding van der Waals forces the lattices were relaxed using the optB86b-vdW functional~\cite{Dion:2004ce}~\footnote{core configuration:$Ta: [Xe] 4f^{14} 5d^3 6s^2 6p^0$ and $S: [Ne] 3s^2 3p^4 3d^{-2}$}. 
For the 2H-${AA^{\prime}}$-phase (space group D46h (P63/mmc)) we obtained relaxed
lattice constants of $a_\text{lat}=3.31 \text{\AA}$ and $c_\text{lat}=12.02 \text{\AA}$, containing two S-Ta-S monolayers seperated by an 
interlayer distances of $\Delta = 2.92 \text{\AA}$ in the unit cell. X-ray diffraction of 2H-TaS$_2$ in powder form yield similar lattice parameters at $a_\text{lat}=3.32 \text{\AA}$ and $c_\text{lat}=12.10 \text{\AA}$ \cite{Jellinek:1962fq}. 
For the 1H monolayer a vacuum of $15 \text{\AA}$ was taken into account to prevent surface-surface hybridization. 

\begin{figure}[th]
\centering
\includegraphics[width=0.98\columnwidth]{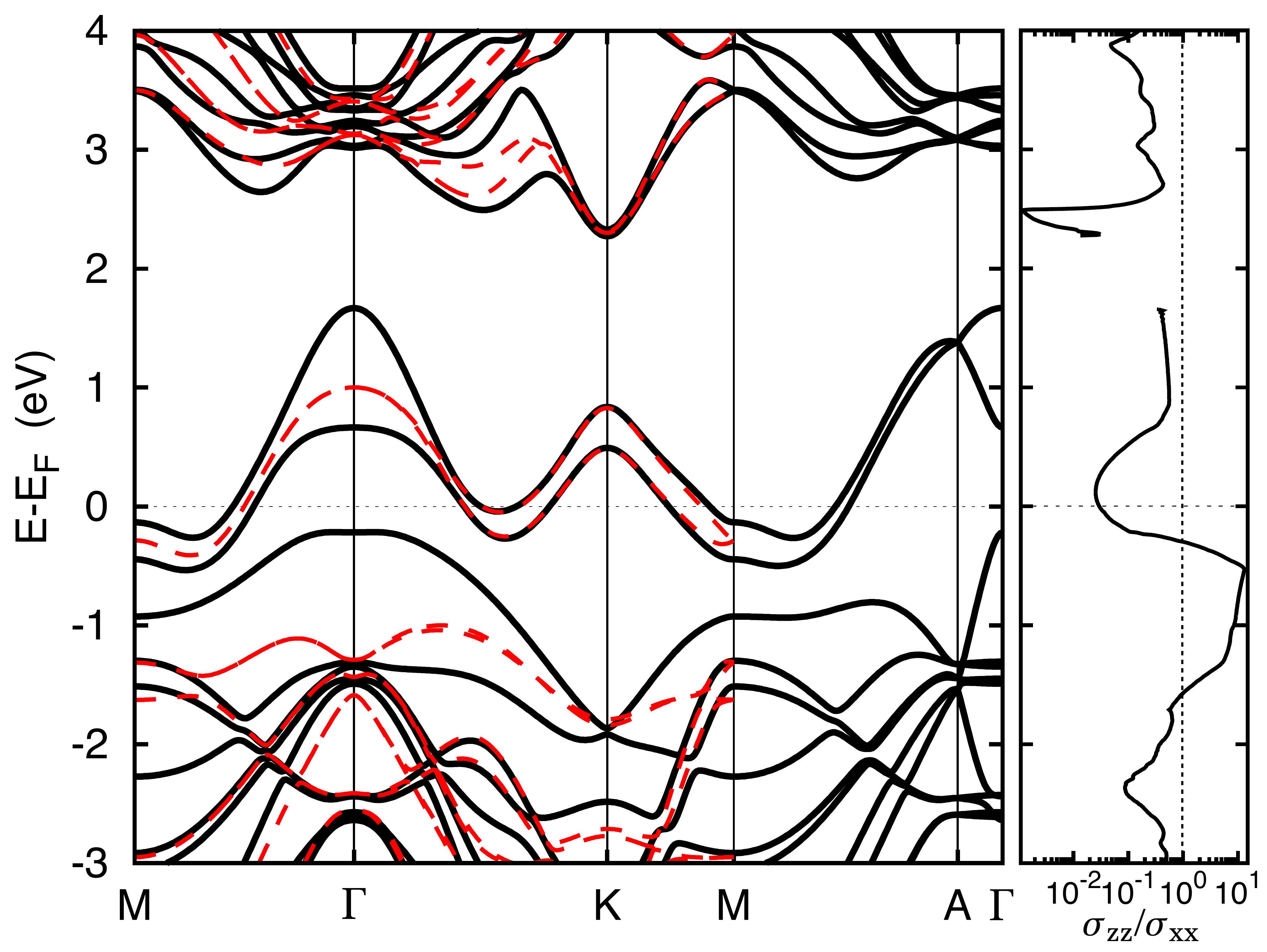}
  \caption{Band structure of bulk 2H-TaS$_2$ (black solid lines) and monolayer 1H-TaS$_2$ (red dashed lines). The corresponding electrical conductivity anisotropy $\nicefrac{\sigma_{\text{zz}}}{\sigma_{\text{xx}}}$ for bulk 2H-TaS$_2$ is shown in the right panel.}
\label{fig1}
\end{figure}

\begin{figure*}[!th]
\centering
\includegraphics[width=0.85\textwidth]{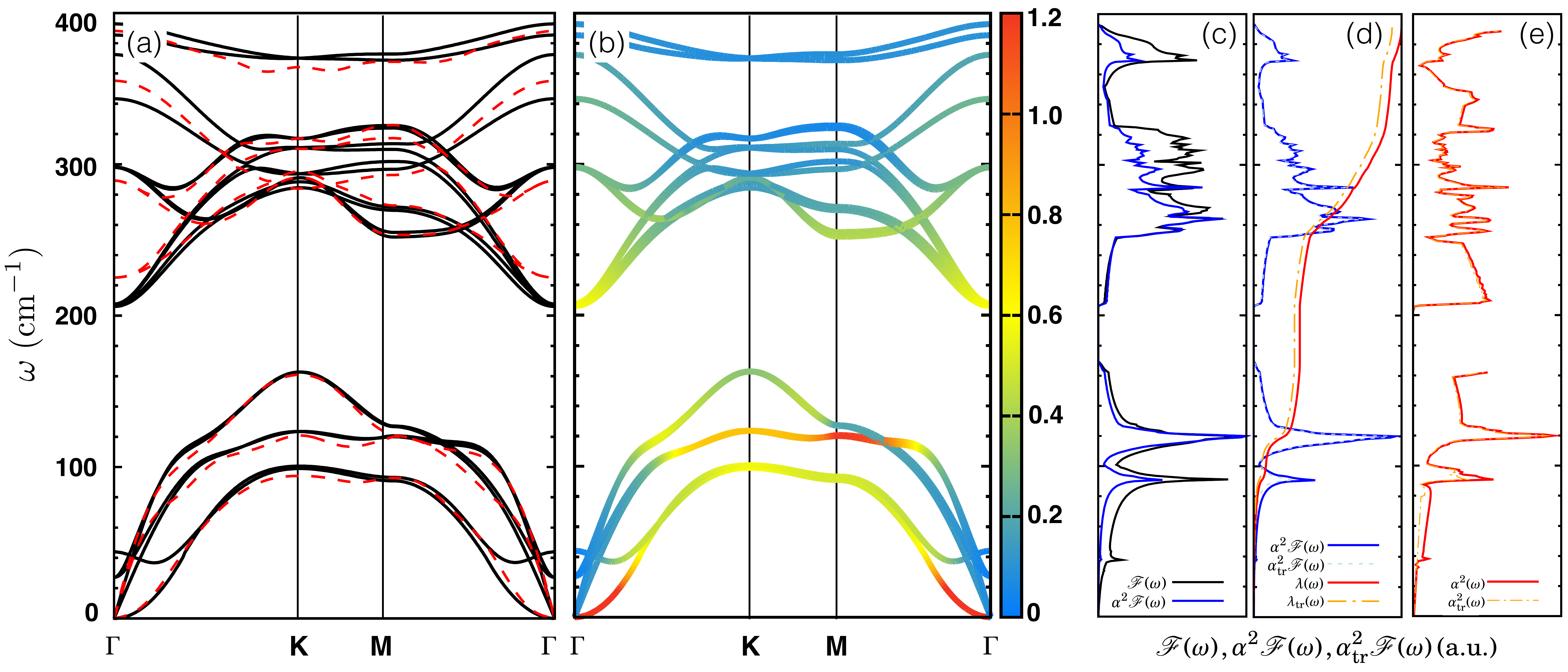}
  \caption{(a) Phonon dispersion of bulk 2H-TaS$_2$ (black solid lines) and monolayer 1H-TaS$_2$ (red dashed lines). (b) Phonon dispersion of bulk 2H-TaS$_2$ . The color code (blue to red; $\lambda=0$ to $\lambda=1.2$) depicts the unitless electron phonon coupling parameter $\lambda_{\bm{q}}$. An unconventionally strong coupling of phonons with $\bm{q} \approx \overline{\Gamma M}$ and $\omega=\unit[120]{cm^{-1}}$ was found. (c), (d), (e) the phonon density of states $\mathcal{F}(\omega)$, the isotropic (transport) \'Eliashberg spectral functions $\alpha^{2}\mathcal{F}_{(\text{tr}})(\omega)$ and the frequency dependent (transport) coupling strength $\alpha^{2}_{(\text{tr}})(\omega)$, respectively. In (d) the cumulative value for the (transport) electron-phonon coupling constant $\lambda_{(\text{tr})}$ is additionally shown as red solid (yellow dashed) line.}
\label{fig2}
\end{figure*}
While being moderately large in comparison to conventional spin-orbit driven materials, e.g. Bi$_2$Te$_3$, spin-orbit coupling (SOC) leads often to visible band splittings in two-dimensional TMDs. The H-phase own a lattice without inversion symmetry and thus the spin degeneracy is fully removed, except for the $\Gamma$-point. 
TaS$_2$ is metallic due to a half-filled $5d_{z^2}$ orbital and these splittings are most pronounced near the Fermi energy in the vicinity of the $K$ high symmetry point, reaching values of around $\unit[350]{meV}$. 
As shown in \F{fig1} the variation in the 
electronic band structures of 1H and 2H-TaS$_2$ are rather small in the vicinity of $E_F$.
Lifted degeneracies in the bulk phase at the $\Gamma$, $M$ and $K$ high symmetry points, as well as a closing of the electronic gap below the Fermi energy in the bulk case are the most distinct effects of the SOC. 
We are aware that more sophisticated methods, e.g. hybrid functionals or $GW$ calculations, lead to an opening of the gaps above and below $E_F$ of around $\unit[0.5]{eV}$. This can change e.g. optical properties and was discussed in an previous study by us \cite{Gjerding:2017ee}. 
However, due to the metallic character of TaS$_2$ and the energetic distance of the lower $p$-like band to $E_F$, these changes do not alter the electronic transport results. This is different for bulk 2H-TaSe$_2$ and further discussed in the supplemental material~\cite{supp}, where band structures based on the HSE06-functional are shown for all four systems, additionally. 

With our main focus being on to the impact of temperature effects onto the electronic transport properties and band renormalization, the 
core of the calculations are the electron-phonon matrix elements, omitting band indices,  
\begin{equation}\label{eq:me}
\me=\sqrt{\frac{\hbar}{2NM\omega_{\q}}}\langle\Psi_{\ko+\q}\left|\partial V_{\q}\right|\Psi_{\ko}\rangle
 \end{equation}
which manifest the interaction of the initial ($\Psi_{\ko}$) and final ($\Psi_{\ko+\q}$) electronic states via a phonon $(\q)$. Here, $\partial V_{\q}$ is the first-order derivative of the Kohn-Sham potential with respect to the atomic displacements induced by a phonon 
mode $\nu$ with frequency $\omega_{\q}$ \cite{Savrasov:1994fq,Savrasov:1996uy}.
The matrix elements were initially obtained on coarse 54x54x18 (54x54x1) k-meshes and 9x9x3 (9x9x1) q-meshes for bulk 2H-TaS$_2$ (monolayer 1H-TaS$_2$), respectively. 
To provide a consistent methodological basis spin-orbit effects were fully taken into account while determining the vibronic properties. We applied an electronic smearing of $\unit[0.03]{Ry}$ during the calculation of 
the vibronic properties to prevent the systems to undergo a charge-density wave (CDW) transition, which for bulk 2H-TaS$_2$ occurs below $\unit[75]{K}$~\cite{Coleman:1992cw,Yan:2015fr}. 

\subsection{Electron-phonon coupling}
The phonon band structures of bulk 2H-TaS$_2$ and its monolayer counterpart are shown in \F{fig2}(a). 
Peculiar for 2D materials are the soft out-of-plane acoustic modes (ZA), or flexural acoustic phonons, presenting a quadratic dispersion at small momenta \textbf{q}. 
In 2H-TaS$_2$, as well as 1H-TaS$_2$, high-frequency optical modes are separated from the
low-frequency modes by a gap of around $\unit[40]{cm^{-1}}$, comparable to other TMDs \cite{MolinaSanchez:2011hb}. 
As expected, bearing in mind the weak interlayer coupling, the phonon branches are almost doubly degenerate for the 2H polytype at frequencies 
larger than $\unit[50]{cm^{-1}}$ and no significant variation between the two material systems could be found. 
The most noticeable deviation can be found for small frequencies, where the bulk branches
split into acoustic branches that approach zero frequency for vanishing \textbf{q} (corresponding to in-phase oscillation of equivalent atoms of adjacent monolayers) 
and \textit{optical} modes that approach a finite value (corresponding to a phase-difference of $\pi$ in the oscillation of adjacent monolayers) \cite{Wirtz:2004eb}.
The total phonon bandwidth ($\unit[390]{cm^{-1}}$) for TaS$_2$ is noticeably smaller than for other prototype 2D materials, e.g. graphene ($\unit[1700]{cm^{-1}}$) or MoS$_2$ ($\unit[500]{cm^{-1}}$), which in turn could lead to strongly reduced lattice thermal conductivity. 

With the knowledge of the electron-phonon matrix elements, the effectiveness of phonons with energy $\hbar \omega$ to scatter electrons 
can be expressed by means of the, \textit{a priori} state-dependent, \'Eliashberg spectral function $\alpha^2 \mathcal{F}_{\mathbf{k}}(E,\omega)$ \cite{Mahan:1990ie,Rittweger:2016vo}
\begin{eqnarray}\label{eq:a2Fk}
\alpha^{2}\mathcal{F}^{\pm}_{\mathbf{k}, (\text{tr})}(E,\omega)=\frac{1}{N_\mathbf{q}}\sum_{\q}\delta(\omega-\omega_{\q})\left|\me\right|^2 \\ \times \delpmE
\times \left(1-\frac{\mathbf{v}_{\kp} \cdot \mathbf{v}_{\mathbf{k}}}{|\mathbf{v}_{\kp}| |\mathbf{v}_{\mathbf{k}}|}\right) \nonumber
\end{eqnarray}
The last term in \eq{eq:a2Fk} accounts for back-scattering effects via the change of velocity during the scattering process, 
which then directly leads to the transport \'Eliashberg spectral function. 
In an iterative solution of the full Boltzmann equation, this term would largely correspond to the the scattering-in term. 
In \F{fig2}(d) the difference between $\alpha^2 \mathcal{F}(\epsilon_F,\omega)$ and its transport counterpart are shown. 
We find both to be functional almost identical, with $\alpha^2 \mathcal{F}_{(\text{tr})}(\epsilon_F,\omega)$ 
being suppressed to around 64\% (96\%) of the value for $\alpha^2 \mathcal{F}(E,\omega)$ for the acoustic (optical) modes. 
While similarity of $\alpha^2 \mathcal{F}(\epsilon_F,\omega)$ and  $\alpha^2 \mathcal{F}_{(\text{tr})}(\epsilon_F,\omega)$ is not a general trend, it has been observed before, experimentally and theoretically, for bulk metals \cite{Savrasov:1996uy,Ponce:2016dv}. 
This trend indicates that forward scattering (small angles between $\mathbf{v}_{\mathbf{k}'}$ and $\mathbf{v}_{\mathbf{k}}$) 
is preferred over backward scattering (large angles between $\mathbf{v}_{\mathbf{k}'}$ and $\mathbf{v}_{\mathbf{k}}$). 

Integration of the isotropic (transport) \'Eliashberg spectral function leads to 
the unitless (transport) electron-phonon coupling constant
${\lambda_{(\text{tr})}=2\int_0^{\infty} d\omega \, \frac{\alpha^2 \mathcal{F}_{(\text{tr})}(\omega)}{\omega}}$
which is shown, resolved for phonon modes $\bm{q}$, in \F{fig2}(d) \footnote{This assumes summing over \textbf{k} instead of \textbf{q} in \Eq{eq:a2Fk}.}. 
The electron phonon coupling spreads values of almost zero to above unity. 
Especially in the low frequency regime, and thus predominantly in the low temperature range, the coupling $\lambda_{\bm{q}}$ is largest for the out-of-plane ZA mode. 
The latter stems mainly from the fact that in the long-wavelength limit $\mathfrak{g}_{\text{\tiny{ZA}}}^2$ is almost constant and thus $\lambda_{\bm{q}}^{\text{\tiny{ZA}}} \propto \bm{q}^{-2}$. For the in-plane acoustic modes we find $\mathfrak{g}_{\text{\tiny{LA,TA}}}^2 \propto \bm{q}$ and hence a constant value $\lambda_{\bm{q}}^{\text{\tiny{LA,TA}}} \approx 0.13$.
Very large values of $\lambda_{\bm{q}} \approx 1.2$ can be found nearby the M high symmetry point at frequencies of $\omega_{\bm{q}} \approx \unit[120]{cm^{-1}}$. 
The latter stems to a large extent from the topology of the Fermi surface yielding a high probability, i.e. an enhanced nesting function, 
dominantly for intra-and interband scattering for initial electronic states $\bm{k}$ into final states $\bm{k}\pm \bm{q}$ with $\bm{q} \approx \overline{\Gamma M}$. We note that this is a  material specific feature and could not be found as pronounced for monolayer 1H-TaSe$_2$~\cite{supp}. 
\begin{figure}[!t]
\centering
\includegraphics[width=0.45\textwidth]{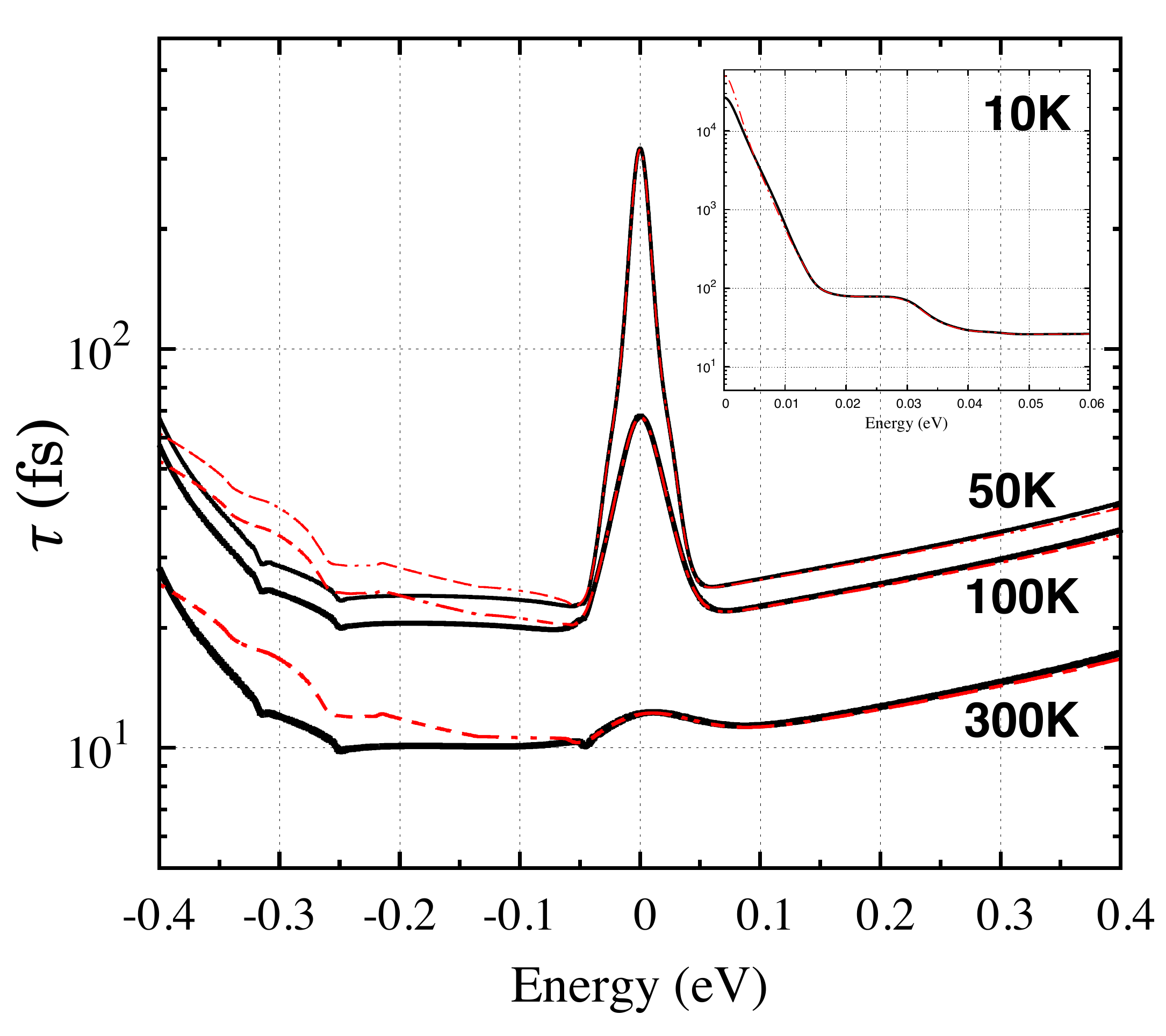}
  \caption{(a) Energy-averaged transport relaxation times with respect to the Fermi energy for monolayer 1H-TaS$_2$ (black solid lines) and bulk 2H-TaS$_2$ (red dash-dotted lines) at three distinct temperatures.}
\label{fig3}
\end{figure}
\begin{figure*}[!t]
\centering
\includegraphics[width=0.85\textwidth]{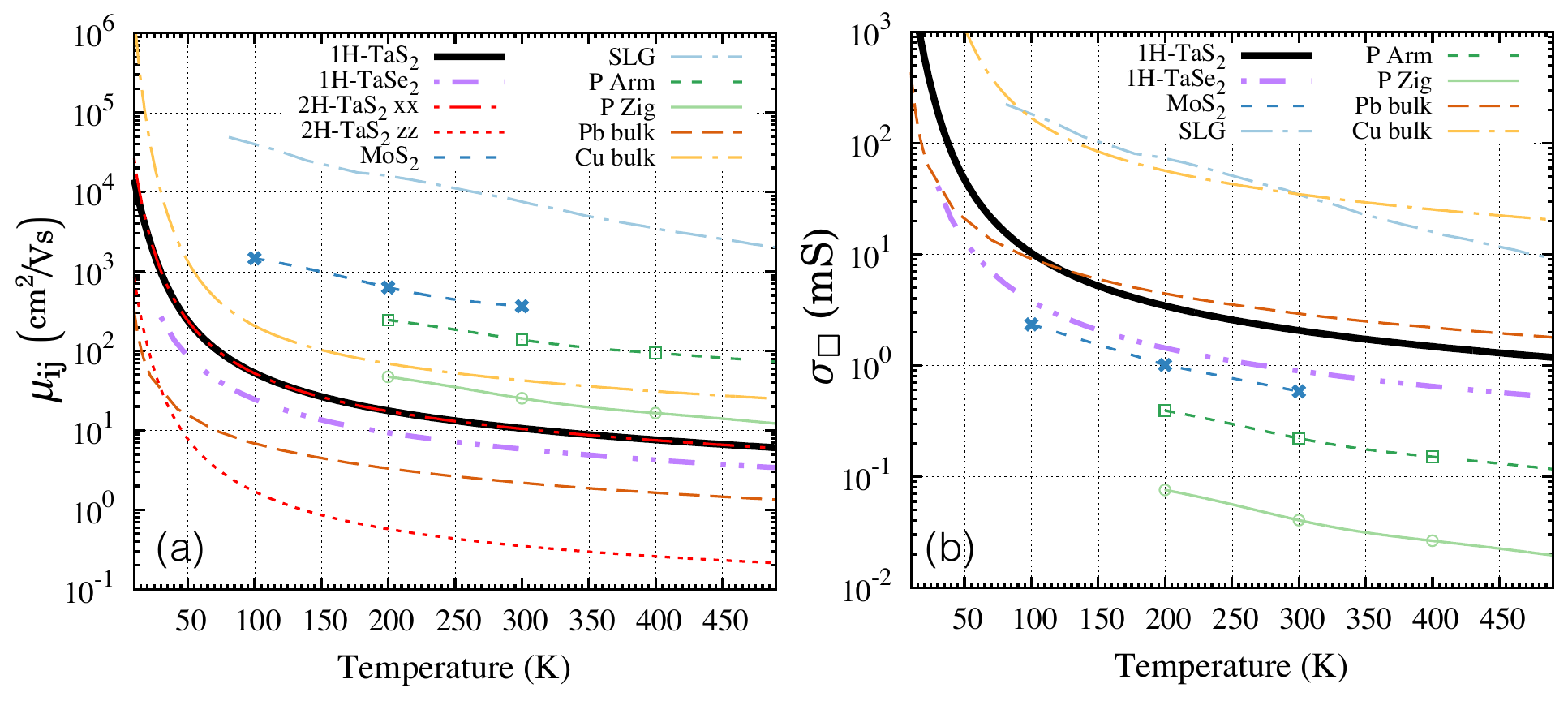}
  \caption{(a) Anisotropic electronic mobilities for monolayer 1H-TaS$_2$ (black solid lines) and bulk 2H-TaS$_2$ (red lines), as well as monolayer 1H-TaSe$_2$ (purple dash-dotted lines) in comparison to 
monolayer MoS$2$ at a charge carrier concentration of $\unit[1\times10^{13}]{cm^{-2}}$ (Ref.~\onlinecite{Gunst:2016ff}), 
graphene at a charge carrier concentration of $\unit[2.86\times10^{13}]{cm^{-2}}$ (Ref.~\onlinecite{Park:2014tf}), monolayer phosphorene at a charge carrier concentration of 
$\unit[1\times10^{13}]{cm^{-2}}$ along the armchair and zig-zag direction (Ref.~\onlinecite{Liao:2015ca}), as well as bulk Pb and bulk Cu. (b) Sheet conductances for the two-dimensional materials. Values for Pb and Cu are presented at a representative thickness of $\frac{c_{\text{2H-TaS}_2}}{2}=6.01 \text{\AA}$.}
\label{fig4}
\end{figure*}
The cumulative electron-phonon coupling parameters, depicted in \F{fig2}(d), were found to be $\lambda=0.38$ and $\lambda_{(\text{tr})}=0.34$. For monolayer 1H-TaSe$_2$ we calculated values of $\lambda=0.50$ and $\lambda_{(\text{tr})}=0.46$. The enhancement of $\lambda$ is most prominently related to an reduced phonon bandwidth and we give further details in the supplemental material~\cite{supp}.
Applying the \textsc{Allen-Dynes} formula~\cite{Allen:1975jg} to estimate the superconducting transition temperature, we obtain $T_{C}=\unit[0.54]{K}$ 
using an Heine-Morel parameter\cite{Morel:1962fe} of $\mu^{\ast}=0.11$ representing the repulsive retarded Coulomb potential for bulk 2H-TaS$_2$ and monolayer 1H-TaS$_2$, respectively. 
While our calculated $T_{C}$ for bulk 2H-TaS$_2$ is in excellent agreement to recent experimental findings \cite{NavarroMoratalla:2016df}, 
the superconducting transition temperature of ultra-thin TaS$_2$ was experimentally found to be larger at $T_{C}=\unit[1.79]{K}$. 
To obtain this temperature a value for the retarted Coulomb parameter of $\mu^{\ast}=0.057$ would have to be assumed~\cite{supp}. 
This would be in contrast to the fact that 2D electronic systems are rather poor at screening interactions and the Coulomb interaction between 
electrons should be much stronger in 2D than in 3D \cite{Olsen:2016js}, i.e. generally $\mu^{\ast}_{2D} > \mu^{\ast}_{3D}$. 

The \'Eliashberg spectral function 
$\alpha^{2}\mathcal{F}_{\mathbf{k}}(E,\omega)$ can be directly linked to the imaginary part $\Sigma^{\prime\prime}$ 
of the electron self energy induced 
by interaction with a phonon in state $\bm{q}$ as 
\begin{align}
\label{eq:lw}
\Sigma^{\prime\prime}_{\ko,(\text{tr})}(E) &= \pi\int_0^{\infty}d\omega \nonumber \\
\{ &\alpha^{2}\mathcal{F}^{-}_{(\text{tr})}(E,\omega) \left(1+b(\omega) +f(E+\hbar\omega\right) + \nonumber \\
&\alpha^{2}\mathcal{F}^{+}_{(\text{tr})}(E,\omega) \left(b(\omega) - f(E-\hbar\omega\right) \}
\end{align}
Here, $b$ and $f$ are the equilibrium Bose-Einstein and Fermi-Dirac distribution functions, respectively and introduce the explicit temperature dependence. 
One gains access to the total complex electron-phonon selfenergy, 
i.e. $\Sigma_{\mathbf{k}}(E,T)=\Sigma_{\mathbf{k}}^\prime(E,T)+\mathfrak{i}\Sigma_{\mathbf{k}}^{\prime\prime}(E,T)$, via a Kramers-Kronig relation \cite{Levy:2000vy}. 
Furthermore we can introduce the spectroscopic electronic linewidth as $\Gamma_{\bm{k}} =2 \Sigma^{\prime\prime}_{\bm{k}}$ 
and the phonon-limited transport lifetime $\tau_{\bm{k}} = -\nicefrac{\hbar}{2 \Sigma^{\prime\prime}_{\ko,(\text{tr})}}$. 
The energy-surface-averaged transport lifetime\footnote{The energy-surface-averaged transport lifetime is defined as \unexpanded{$\langle \tau_{\bm{k}} \rangle = \nicefrac{\sum\limits_{k} \tau_{\bm{k}} \delta(\E- \epsilon)}{\sum\limits_{k} \delta(\E- \epsilon)}$}. For purposes of presentation shown results are afterwards smoothed by a spline function.} is depicted in \F{fig3} together for bulk 2H and monolayer 1H-TaS$_2$ at three distinct temperatures. 
Again, due to the similarities in the electronic and vibronic structure, no striking difference between the lifetimes of bulk and monolayer TaS$_2$ can be stated. 
\footnote{We note that the overall energy dependency of $\tau(E)$ in this particular case is functionally best described by the available iso-energy area $\oint_{\E=E} \text{d}\bm{k} \frac{\text{d}E}{\hbar \bm{v}_{\bm{k}}}$. However, a qualitative argumentation using the the electronic density of states still holds.}. The pronounced peak of the carrier lifetime around the Fermi energy can be related to a freeze-out of phonons and the narrowing of the Fermi surface resulting in a limited phase space for phonon scattering \cite{Kaasbjerg:2012wl}. 
Technically, this effect is represented in \Eq{eq:lw} by weighting the electron-phonon matrix elements with the 
Bose-Einstein and Fermi-Dirac distribution functions $\mathcal{H}(b,f)$. The peak itself can be decomposed into phonon absorption and emission processes relative to energies below and above the Fermi energy, respectively. The lifetimes related to phonon absorption rapidly increase as the electron energy approaches the Fermi energy from below. 
Pronounced shoulders in the lifetimes at around $\unit[25]{meV}$ at very low temperatures of 10K (c.f. inset \F{fig3}) indicate the onset of scattering by optical phonons. Due to the smearing of the Fermi-Dirac distribution and the increasing availability of short-wavelength phonons the lifetime peak steadily decreases at higher temperatures \footnote{The reader should keep in mind that 2H-TaS$_2$ undergoes an incommensurate phase transition at $T \approx \unit[75]{K}$, which is not covered within our methodology \cite{Coleman:1992cw}.}.


\subsection{Transport properties}
On the basis of the transport relaxation times, the electronic transport properties are subsequently obtained by solving a linearized Boltzmann equation in relaxation time approximation (RTA) under the explicit influence of electron-phonon scattering as introduced in Refs. \onlinecite{Mertig:1999br,Hinsche:2015er,Zhou:2016tt,Rittweger:2016vo}.
The components of the electrical conductivity tensor can be stated as 
\begin{eqnarray}
\sigma_{ij} = \frac{e^2}{V_{\tiny{\text{uc}}}N_{\bf{k}}} \sum_{\mathbf{k}} \tau_{\mathbf{k}} \mathbf{v}_{\mathbf{k}}^i \mathbf{v}_{\mathbf{k}}^j  \left( -\frac{\partial f_{\mathbf{k}}}{\partial \epsilon_{\mathbf{k}}} \right) 
\label{eq:mobility}
\end{eqnarray}
where $V_{\tiny{\text{uc}}}$ is the unit cell volume and $N_{\bf{k}}$ is the number of initial electronic states in the reciprocal space. One obtains the sheet conductance for a film of thickness d directly as $\sigma_{\square}= \frac{d}{2} (\sigma_{xx} + \sigma_{yy})$ and the unipolar phonon limited electron mobility tensor at a given charge carrier concentration n as $\mu_{ij} = \nicefrac{\sigma_{ij}}{n e}$ .
The convergence of the sums in \Eq{eq:a2Fk} and \Eq{eq:mobility} is ensured by employing adaptive tetrahedron schemes \cite{Zahn:2011ci,Kawamura:2014cr,Rittweger:2016vo} and using, depending on the topology of the iso-energetic surface, 
a few $10^4$ ($10^3$) $\bm{k}$-points and $10^5$ ($10^5$) $\bm{q}$-points for bulk and monolayer, respectively \cite{supp}. 
In \F{fig4}(a) we show the electronic mobilities for bulk and monolayer TaS$_2$, as well as monolayer TaSe$_2$, in comparison to other two-dimensional materials and well studied bulk systems. 
The in-plane mobilities of TaS$_2$ are found to be $\mu=\unit[10.5]{\frac{cm^2}{Vs}}$ at $\unit[300]{K}$. Thus TaS$_2$ classifies itself as a good metal, 
right in between bulk Cu ($\mu=\unit[43]{\frac{cm^2}{Vs}}$) and Pb ($\mu=\unit[2.2]{\frac{cm^2}{Vs}}$). 
These values are lower or comparable to those of heavily doped~\footnote{Technically the doping is achieved, by a rigid shift of the chemical potential into the bands of the material. Experimentally electrical gating would most closely correspond to this situation.} two-dimensional materials, e.g. MoS$_2$, phosphorene or graphene. 
The mobilities decrease as expected at rising temperature, when the phonon phase-space increases and enlarges the probability of electron-phonon scattering. 
At high temperatures, where the phonon modes contribute almost equally to the  scattering, the semi-classical limit $\mu \propto T^{-1}$ is achieved \cite{Fabian:1999cm}. The in-plane mobility of monolayer 1H-TaSe$_2$ was found to be $\mu=\unit[5.8]{\frac{cm^2}{Vs}}$ at $\unit[300]{K}$, almost a factor two lower than its isoelectronic counterpart 1H-TaS$_2$. One reason is the enhanced electron-phonon coupling and hence reduced electron-phonon relaxation times. Second to that we also found the Fermi-surface averaged velocities to be around 17\% smaller in 1H-TaSe$_2$ compared to 1H-TaS$_2$. We show more details on the electronic, vibronic and el-ph properties of 1H-TaSe$_2$ the supplemental material \cite{supp}, where conclusions can be drawn  in a similar way as done here for monolayer 1H-TaS$_2$.

As expected almost the entire temperature-dependence of $\mu$ arises from the temperature-dependence of the relaxation times introduced via \Eq{eq:lw}. The remaining band structure kernel in \Eq{eq:mobility} varies only a few percent over the whole temperature range, mainly linked to a change of the temperature-dependent chemical potential.

We also highlight the out-of plane mobility for bulk 2H-TaS$_2$, noting $\mu=\unit[0.35]{\frac{cm^2}{Vs}}$ at $\unit[300]{K}$ 
(c.f. red dotted line in \F{fig4}(a)). 
The reduction stems dominantly from heavily reduced out-of plane carrier velocities in the vicinity of the Fermi energy. 
As shown by the conductivity anisotropy in the right panel of \F{fig1}, out-of-plane mobilities very close to in-plane mobility values 
can be achieved under heavy electron doping. This scenario would assume almost filling up the half-filled $5d_{z^2}$ band, 
which seems to be unfeasible to achieve by chemical doping. However an inherent pseudo-doping of about 0.3 electrons was recently 
observed for epitaxially grown monolayer 1H-TaS$_2$~\cite{Sanders:2016gz,Wehling:2016wd}. 
Accessing high out-of-plane mobilities via electrical field gating turns out to be very challenging. Assuming a SrTiO$_3$ substrate of $\unit[100]{nm}$ thickness and 
a film thickness of $\unit[10]{nm}$ 2H-TaS$_2$ would still require a gating voltage in the order of $\unit[100]{V}$ to gain values close to $\mu_{||} \approx \mu_\perp$. 
We note, that $\mu_{\perp} > \mu_{||}$ can be achieved for $\unit[-1.5]{eV} < E-E_F < \unit[-0.3]{eV}$, which originates largely from $p$-orbitals forming highly dispersive bands with high out-of-plane velocities. As mentioned before and shown in the supplemental material, applying GW or hybrid functional calculations, these bands will be shifted downwards in energy by about $\unit[0.6]{eV}$, making them unlikely to accessed by hole doping \cite{Gjerding:2017ee,supp}. 

As noted earlier, with regard to potential application as transparent conductive electrodes, the electrical sheet conductance 
should be the material property in focus and is shown in \F{fig4}(b). 
TaS$_2$ clearly outperforms the conventional heavily doped semi-conducting 2D materials 
MoS$2$ and Phosphorene and decreases the gap to graphene by showing values of $\sigma_{\tiny{\square}}=\unit[2]{mS}$ at room temperature. 
With this still being an order of magnitude worse than the sheet conductances of heavily doped Graphene or a comparable layer of Cu, the advantages of 
thin film TaS$_2$ are noteworthy. On the one hand, in contrast to Graphene, TaS$_2$ is an inherent metal and thus does not require to be doped or gated 
to achieve a feasible sheet conductances. On the other hand it can easily be exfoliated due to its weak interlayer interactions and thinned down 
to atomic thicknesses, bearing a huge advantage to traditional metals, e.g. Cu or Pb. 
These material properties, together with a high optical transmittance in the visible range, 
enables H-TaS$_2$ to be a promising system for application as a transparent conducting oxide \cite{Gjerding:2017ee}.

\begin{figure*}[t]
\centering
\includegraphics[width=0.8\textwidth]{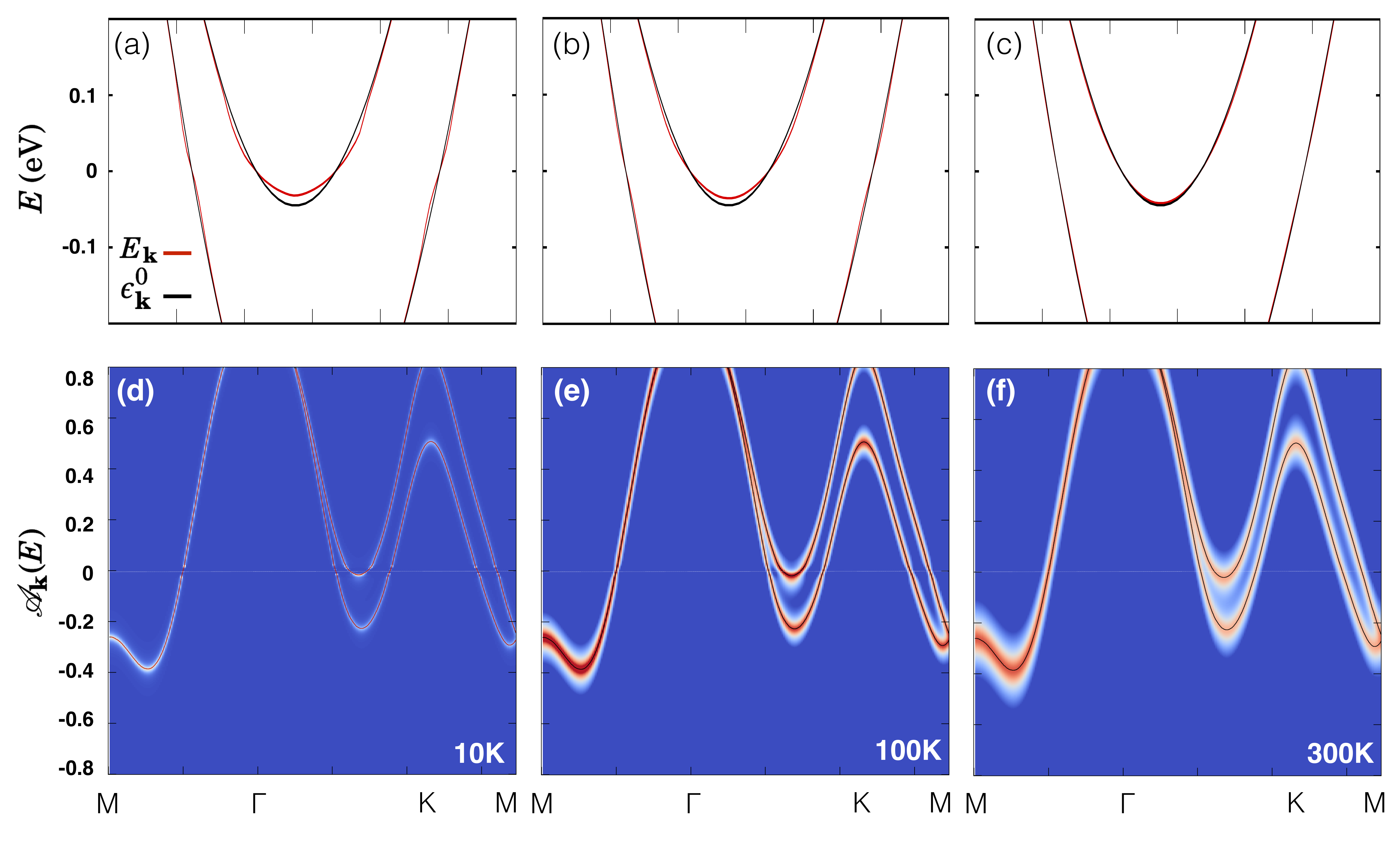}
\caption{The electronic spectral function $\mathcal{A}_{\mathbf{k}}(E)$ with respect to the Fermi energy for monolayer 1H-TaS$_2$, (d)-(f) incorporating electron-phonon interaction as calculated via \Eq{eq:specfun} at different temperatures (color scale: blue-white-red; low to high intensity. Thin black lines show the maximum highlighted in the panel above). (a)-(c) zoom-in of the corresponding extracted maximum of $\mathcal{A}_{\mathbf{k}}(E)$ around $E_F$ on the $\Gamma-K$ path (red lines). The black lines indicate the bare band structure without interaction\footnote{Note that the spectral functions have been obtained in a structural phase not possessing a charge-density wave.}}
\label{fig5}
\end{figure*}

\subsection{Phonon induced band renormalization}
Having discussed the single-particle electronic and vibronic properties, as well as the phonon limited electronic transport properties of H-TaS$_2$, we focus in the following 
on the renormalization of the electronic structure due to electron-phonon-interaction. 
Assuming the photoemission intensity is proportional to the hole spectral function of the sample \cite{Hofmann:2009kw,Hinsche:2017ej}, the electron spectral function 
\begin{eqnarray}
\mathcal{A}_{\mathbf{k}}(E) =   \frac{1}{\pi}\frac{\left| \Sigma_{\mathbf{k}}^{''}(E) \right|}{\left[E - \epsilon_{\mathbf{k}}^{0}-\Sigma_{\mathbf{k}}^{'}(E)\right]^2 + \left[ \Sigma_{\mathbf{k}}^{''}(E) \right]^2}.
\label{eq:specfun}
\end{eqnarray}
can be used to describe the electronic structure of a solid in the presence of many-body effects. 
We note, that within our framework the uncorrected self-energy corresponds to the Fan contribution \cite{Fan:1950cb} only, which is a resonable approach for metallic systems \cite{Giustino:2017ge}. $\Sigma_{\mathbf{k}}^{'}(E=E_F)=0$ is enforced, such that the quasiparticle Fermi level fulfills Luttinger’s theorem of Fermi surface volume conservation \cite{Luttinger:1960ht}.

Comparable to angle-resolved photoemission (ARPES), $\mathcal{A}_{\mathbf{k}}(E)$ can be viewed 
as the probability of finding an electron with energy E and momentum $\mathbf{k}$ at a given temperature T \cite{Zhou:2007ec,Hofmann:2009kw}. 
In \F{fig5}(d)-(f) the electronic spectral function around the Fermi energy is shown for monolayer 1H-TaS$2$ at different temperatures. 
Obviously the overall linewidth broadening, i.e. $\Gamma_{\bm{k}} =2 \Sigma^{\prime\prime}_{\bm{k}}$, 
increases with increasing temperature. 
The linewidth broadening spreads Fermi surface-averaged values of $\Gamma=\unit[0.025, 9.7, 54]{meV}$ at $T=\unit[10, 100, 300]{K}$. 
Overall, a noticeable $\bm{k}$-anisotropy of the linewidths was found, 
with values of $\Gamma_{\bm{k}}$ varying almost a factor 3 throughout the Brillouin-zone~\cite{supp}. 
While the imaginary part of $\Sigma$ is related to the spectroscopic linewidth broadening, the real part $\Sigma^{\prime}$ will enable us to analyze the 
electron-phonon renormalized band dispersion $E_{\mathbf{k}}$. For weak to modest electron-phonon interaction, recovering the Rayleigh-Schrödinger perturbation theory \cite{Allen:1967dh}, the maximum of the electronic spectral function can 
be estimated to lowest order as $E_{\mathbf{k}} \approx \E+\Sigma^{\prime}(\E,T)$ and is shown in \mbox{\F{fig5}(a)-(c)} (red lines) in comparison to the 
bare non-interacting band structure $\E$ (black lines) \cite{Giustino:2017ge}. In the low temperature regime (\F{fig5}(a)) clear kinks at energies of about $\pm \unit[45]{meV}$ around $E_F$ 
can be found in the outlying band, roughly corresponding to maximal available phonon energy. 
For the inner band the scenario slightly differs. While kinks are visible for energies above $E_F$ 
no states are available for high frequency phonon absorption below $E_F$. As consequence the local band bottom is energetically lifted and the effective mass increases. 
At $E_F$ $\Sigma^{\prime}=0$ and the volume of the interacting and non-interacting Fermi surface is conserved due to Luttinger’s theorem\cite{Luttinger:1960ht}. 
Similar findings have been observed in the f-electron material USb$_2$, whereas renormalization due to electron-electron interaction was assumed there \cite{Durakiewicz:2010wa}.
At elevating temperatures ((\F{fig5}(b),(c))) the pronounced features in the real part of $\Sigma$ decrease and the bare band structure is almost recovered at room temperature.


\subsection{Conclusion}
On basis of density functional perturbation theory and semi-classical Boltzmann transport, we found 
promising room temperature mobilities and sheet conductances in metallic bulk 2H-TaS$_2$ and its exfoliated monolayer 1H-TaS$_2$. 
The sheet conductances can compete with other established 2D materials and theoretically predict 
TaS$_2$ as a promising material system to be used as a transparent conductor or as atomically thin interconnect.
We furthermore found no significant change of the electron-phonon coupling due to interlayer coupling when stacking 1H monolayers to the bulk 2H-TaS$_2$ polytype.
Hence no phonon-mediated enhancement of the superconducting transition temperature in the thin film compared to the bulk state could be found. To extent and support our findings on the transition metal dichalcogenide TaS$_2$, we discussed our results at key points to comparable findings on the isoelectronic counterpart TaSe$_2$. To keep the the manuscript compact and intelligible, more detailed results on H-TaSe$_2$ can be found in the supplemental material.\\

\textit{Note by the authors:} Shortly after acceptance of this work, we became aware of a very recent experimental study by Wijayaratne \textit{et al.}\cite{Wijayaratne:2017bz} on moment-dependent electron-phonon interaction in bulk 2H-TaS$_2$. The observed anisotropy and total values of the coupling parameter $\lambda$, as well as phonon-induced band renormalizations compare very well to our theoretical findings for TaS$_2$.

\begin{acknowledgments}
NFH received funding within the H.C. \O rsted Programme from the European Union's Seventh Framework Programme and Horizon 2020 Research and Innovation Programme under Marie Sklodowska-Curie Actions grant no. 609405 (FP7) and 713683 (H2020).
\end{acknowledgments}


\section*{References}
\bibliography{paper.bbl}

\end{document}